\documentclass[]{elsarticle}
\makeatletter
\def\ps@pprintTitle{%
 \let\@oddhead\@empty
 \let\@evenhead\@empty
 \def\@oddfoot{}%
 \let\@evenfoot\@oddfoot}
\makeatother
\usepackage{fancyhdr}

\usepackage{epsfig}
\usepackage{amsfonts}
\usepackage{amssymb, graphics, amsmath}
\usepackage{dsfont}
\usepackage{hyperref}
\usepackage{color}
\usepackage{pdflscape}
\usepackage{pifont}
\usepackage{graphicx} 
\usepackage{bm}

\def\slashchar#1{\setbox0=\hbox{$#1$}           
  \dimen0=\wd0                                    
  \setbox1=\hbox{/} \dimen1=\wd1                  
  \ifdim\dimen0>\dimen1                           
    \rlap{\hbox to \dimen0{\hfil/\hfil}}            
    #1                                             
  \else                                          
    \rlap{\hbox to \dimen1{\hfil$#1$\hfil}}        
    /                                           
 \fi}

\newcommand{\tb}{\bar \theta}

\newcommand{\mpi}{m_{\pi}}

\newcommand{\dslash}[1]{#1 \llap{/\kern-0.5pt}}
\newcommand{\Dslash}[1]{#1 \llap{/\kern+1.5pt}}
\newcommand{\DDslash}[1]{#1 \llap{/\kern+2.3pt}}
\newcommand{\dslashh}[1]{#1 \llap{/\kern+1pt}}

\newcommand{\boldtau}{\mbox{\boldmath $\tau$}}
\newcommand{\boldpi}{\mbox{\boldmath $\pi$}}

\newcommand{\bea}{\begin{eqnarray}}
\newcommand{\eea}{\end{eqnarray}}
\newcommand{\bma}{\begin{pmatrix}}
\newcommand{\ema}{\end{pmatrix}}
\newcommand{\nn}{\nonumber}

\makeatletter
\renewcommand*\env@matrix[1][*\c@MaxMatrixCols c]{%
  \hskip -\arraycolsep
  \let\@ifnextchar\new@ifnextchar
  \array{#1}}
\makeatother

\begin{document}
\rhead{LA-UR-16-29007, NIKHEF 2016-057}

\title{Lattice QCD spectroscopy for hadronic CP violation}

\author[ntg]{Jordy de Vries}
\author[lanl]{Emanuele Mereghetti}
\author[inpac]{Chien-Yeah Seng}
\author[lbnl]{Andr{\'e} Walker-Loud}

\address[ntg]{Nikhef, Theory Group, Science Park 105, 1098 XG, Amsterdam, The Netherlands}
\address[lanl]{Theoretical Division, Los Alamos National Laboratory, Los Alamos, NM 87545, USA}
\address[inpac]{INPAC, Department of Physics and Astronomy, Shanghai Jiao Tong University, Shanghai 200240, Shanghai, China}
\address[lbnl]{Nuclear Science Division, Lawrence Berkeley National Laboratory, Berkeley, CA 94720, USA}

\begin{abstract}
The interpretation of nuclear electric dipole moment (EDM) experiments is clouded by large theoretical uncertainties associated with nonperturbative matrix elements. In various beyond-the-Standard Model scenarios nuclear and diamagnetic atomic EDMs are expected to be dominated by CP-violating pion-nucleon interactions that arise from quark chromo-electric dipole moments. The corresponding CP-violating pion-nucleon coupling strengths are, however, poorly known. 

In this work we propose a strategy to calculate these couplings by using spectroscopic lattice QCD techniques. Instead of directly calculating the pion-nucleon coupling constants, a challenging task, we use chiral symmetry relations that link the pion-nucleon couplings  to nucleon sigma terms and mass splittings that are significantly easier to calculate. 
In this work, we show that these relations are reliable up to next-to-next-to-leading order in the chiral expansion in both $SU(2)$ and $SU(3)$ chiral perturbation theory. 
We conclude with a brief discussion about practical details regarding the required lattice QCD calculations  and the phenomenological impact
of an improved understanding of CP-violating matrix elements.

\end{abstract}

\maketitle
\thispagestyle{fancy}

%
\section{Introduction}
The search for phenomena that can explain the apparent shortcomings of the Standard Model (SM) takes place over a large range of energy scales. The LHC explores the high-energy frontier, so far without finding any deviations from SM predictions, whereas low-energy experiments aim to uncover new physics by comparing high-precision measurements to high-precision theoretical predictions. Several classes of low-energy experiments have the potential to probe energy scales comparable to, or even above, the TeV scale probed by the LHC. Examples of such experiments are proton-decay searches, the muon $g\!-\!2$, neutron-antineutron oscillations, rare decays, and searches for permanent electric dipole moment (EDMs). 

The search for EDMs is a particularly active field of research. Very strong bounds on the EDMs of different systems, from the neutron~\cite{Baker:2006ts}, to diamagnetic atoms such as $^{199}$Hg~\cite{Graner:2016ses},
to polar molecules such as ThO \cite{Baron:2013eja}, exist, and are projected to improve by one to two orders of magnitude in the near future. Experiments with other heavy diamagnetic systems, such as $^{225}$Ra \cite{Bishof:2016uqx} and $^{129}$Xe \cite{Xebound} have already set constraints and strong improvements are expected. Furthermore, exciting progress has been made on the proposal to investigate the EDMs of light nuclei in storage rings \cite{Eversmann:2015jnk,Guidoboni:2016bdn}. While a single measurement in any of these systems would only indicate a so far unmeasured source of CP violation, measurements of different, complementary systems could point towards the microscopic source \cite{Lebedev:2004va,deVries:2011an,Dekens:2014jka}.

The interpretation of various EDM experiments relies on the knowledge of nonperturbative 
matrix elements that link operators at the fundamental quark-gluon level to hadronic quantities. An outstanding challenge is the calculation of the nucleon EDMs in terms of CP-violating (CPV) sources in the SM (the QCD $\bar\theta$ term) and beyond. The latter can be categorized in an effective field theory (EFT) picture, 
assuming the new physics is heavy, where the most relevant higher-dimensional operators are
the quark electric dipole moments (qEDMs), chromo-electric dipole moments (qCEDMs), the Weinberg three-gluon operator, and several four-quark operators.
The last year has seen great progress in lattice QCD (LQCD) calculations of the nucleon EDMs  in terms of the $\bar\theta$ term \cite{Shintani:2015vsx,Shindler:2015aqa,Guo:2015tla} and qEDMs \cite{Bhattacharya:2015esa,Bhattacharya:2015wna}, while lattice calculations for the qCEDMs \cite{Bhattacharya:2016oqm} and the Weinberg operator \cite{Shindler_private} have been initiated. 

The interpretation of EDM experiments involving more than one nucleon additionally depends on CPV nucleon-nucleon interactions. The chiral power counting predicts that these mainly depend on one-pion-exchange contributions involving CPV pion-nucleon vertices \cite{Maekawa:2011vs,deVries:2012ab,Bsaisou:2014oka}. The resulting multi-nucleon contributions to nuclear and atomic EDMs often dominate the contributions from the EDMs of the constituent nucleons \cite{Flambaum:1984fb}. Calculations of the CPV pion-nucleon coupling constants are therefore as important as those of nucleon EDMs. In case of the $\bar\theta$ term, chiral-symmetry considerations can be used to connect the leading isoscalar CPV pion-nucleon coupling, $\bar g_0$, to various combinations of octet baryon masses \cite{Crewther:1979pi}. In Ref.~\cite{deVries:2015una}
it was demonstrated that the relation between $\bar g_0$ and the neutron-proton mass splitting is free from large $SU(3)$-flavor breaking corrections through next-to-next-to-leading order (N${}^2$LO) in the chiral expansion, whereas the symmetry relations between the other $SU(3)$ flavor-octet mass splittings and $\bar g_0$ suffer from large $SU(3)$ breaking corrections.
Using an average~\cite{Walker-Loud:2014iea} of
state-of-the-art lattice calculations of the isovector nucleon mass splitting~\cite{Beane:2006fk,Blum:2010ym,Horsley:2012fw,deDivitiis:2013xla,Borsanyi:2013lga,Borsanyi:2014jba}, 
a determination of $\bar g_0(\theta)$ with $\mathcal O(15\%)$ was possible~\cite{deVries:2015una}.

The success of chiral symmetry consideration in the case of the $\bar\theta$ term 
motivates the study of similar relations for the qCEDMs. 
In this case the CPV pion-nucleon couplings can be linked to hadron masses and mass splittings induced by CP-conserving quark chromo-magnetic dipole moments (qCMDMs) \cite{Pospelov:2005pr,deVries:2012ab,Mereghetti:2015rra}. While these matrix elements are not known, they are very suitable for lattice QCD (LQCD) calculations as they can be performed with simple spectroscopic methods.
A direct LQCD calculation of the CPV pion-nucleon coupling, or the full nucleon EDM resulting from such qCEDM and qCMDM operators is substantially more difficult.
Recent work \cite{Seng:2016pfd}, however, cast doubts on the reliability of this method as higher-order chiral corrections strongly violate the relations.

In this work we demonstrate in some detail how these problems can be avoided by a suitable modification of the strategy.
We show that relations for $\bar g_0$ and $\bar g_1$ can be written down that are protected from all next-to-leading-order (NLO) and the bulk of the N${}^2$LO corrections, paving the way for an accurate extraction of the CPV pion-nucleon couplings from LQCD. These relations are provided in Eq.~\eqref{eq:g0g1_mod}.

%
\section{CP violating interactions and hadron spectroscopy}

%
\subsection{CP violation at the quark-gluon level}
As discussed above, in an EFT approach the relevant CPV interactions involving quarks and gluons consist of the QCD $\tb$ term and several higher-dimensional operators. 
In this work we focus on the qCEDMs, and their chiral partners, the CP-conserving qCMDMs. In its most general form, the Lagrangian is then given by
\begin{equation}\label{eq:q_Lcpv}
\mathcal L_{\textrm{QCD}} = 
	\bar{q} i \Dslash{D} q 
	- \bar q \mathcal M q 
	+ \bar q i \gamma_5 q\,m_{*} (\bar \theta - \bar \theta_{\mathrm{ind}}) 
	+ r\,\bar q i \gamma_5 \tilde d_{CE} q
	-\frac{g_s}{2} \bar q \sigma^{\mu\nu} G_{\mu\nu}(\tilde{d}_{CM} + \tilde{d}_{CE} i \gamma_5)  q\ ,
\end{equation}
where $q$ is the quark field $q = (u,d,s)$, $G_{\mu\nu} = G^a_{\mu\nu} t^a$ the gluon field strength contracted with $SU(3)$ generators, $\mathcal M$ the quark mass matrix, $\mathcal M = \textrm{diag}(m_u,m_d,m_s)$, $\bar\theta$ the QCD $\bar\theta$ term, and $\tilde d_{CE}$
and $\tilde d_{CM}$ contain the quark chromo-electric and chromo-magnetic couplings, $\tilde d_{CE} = \textrm{diag}(\tilde d_u, \tilde d_d, \tilde d_s)$ and 
$\tilde d_{CM} = \textrm{diag}( \tilde c_u, \tilde c_d, \tilde c_s)$. 
To write the Lagrangian \eqref{eq:q_Lcpv} we applied the anomalous $U(1)_A$ symmetry to rotate the QCD $\bar\theta$ term into a complex mass term and performed further non-anomalous axial $SU(n_f)$ rotations to align the vacuum of the theory with and without CP violation \cite{deVries:2012ab,Bsaisou:2014oka,Dashen:1970et,Baluni:1978rf}. 
After vacuum alignment, the QCD $\bar\theta$ term is purely isoscalar and proportional to the reduced quark mass 
\begin{equation}
m_* = \left(\frac{1}{m_u} + \frac{1}{m_d} + \frac{1}{m_s}\right)^{-1}  = \frac{\bar m (1-\varepsilon^2)}{2} \left( 1 + \frac{\bar m (1-\varepsilon^2)}{2 m_s} \right)^{-1},
\end{equation}
where we have introduced the notation $\bar m = (m_u + m_d)/2$ and $2 \bar m \varepsilon = m_d - m_u$. Similarly, we introduce $\tilde x_{0,3} = (\tilde x_u \pm \tilde x_d)/2$ for $x \in \{c,d\}$. 
The $SU(n_f)_V$-breaking components of the qCEDM, $\tilde d_3$ and $\tilde d_8 = \tilde d_s - \tilde d_0$, cause vacuum misalignment, which manifests in tadpole couplings of the neutral pion and $\eta$ meson to the vacuum. 
In Eq. \eqref{eq:q_Lcpv} we aligned the theory at the quark-gluon level (see Ref. \cite{Bhattacharya:2015rsa}) which causes the appearance of two corrections to the complex mass term, proportional to the nonperturbative matrix element $r$ and the induced $\bar\theta$ term $\bar\theta_{\mathrm{ind}}$, with
\begin{equation}\label{eq:1.2}
r = \frac{1}{2} \frac{ \langle 0 | \bar q  g_s \sigma_{\mu \nu}   \, G^{\mu \nu} q  | 0 \rangle}{\langle 0 | \bar q q | 0 \rangle}, \quad \bar\theta_{\mathrm{ind}} =  r \textrm{Tr} \left(\mathcal M^{-1} \tilde d_{CE}\right) = r \left( \frac{\tilde d_u}{m_u}+\frac{\tilde d_d}{m_d}+\frac{\tilde d_s}{m_s}\right)\ .
\end{equation}
Alternatively, the alignment can be performed at the level of the hadronic effective field theory \cite{Mereghetti:2010tp,deVries:2012ab}.
The ratio of vacuum condensates $r$ is a dimensionful parameter, and
by naive dimensional analysis (NDA) $r \sim \mathcal O(\Lambda_\chi^2)$, where $\Lambda_\chi \sim 1$ GeV is a typical hadronic scale.
This estimate is in reasonable agreement with a QCD sum rules result $r \sim 0.4$ GeV$^2$ \cite{Belyaev:1982cd}.
The name $\bar\theta_{\mathrm{ind}}$  for the combination of parameters in Eq. \eqref{eq:1.2} is inspired by the Peccei-Quinn (PQ) mechanism  \cite{Peccei:1977hh}.
In the PQ mechanism, if CP violation arises solely from the quark mass term, the theta term dynamically relaxes to 0 and all CP violation is eliminated. In the presence of other sources of CP violation, the PQ mechanism causes $\bar\theta$ to relax to a non-zero value $\bar\theta_{\textrm{ind}}$, proportional to the couplings of the higher-dimensional CPV sources. In the case of the qCEDM, $\bar\theta$ relaxes to the particular combination in Eq. \eqref{eq:1.2} \cite{Pospelov:2005pr,Bhattacharya:2015rsa}. 
We stress that Eq. \eqref{eq:q_Lcpv} is valid regardless of the presence of the PQ mechanism, whose only consequence would be to set  $\bar\theta = \bar\theta_{\mathrm{ind}}$ such that the third term in Eq.~\eqref{eq:q_Lcpv}  disappears.

%
\subsection{CP violation in $SU(2)$ chiral perturbation theory}
We start our discussion by only considering the two lightest quark flavors. We are interested in the effects of the up and down qCEDMs on the interactions among the lightest mesons and baryons: pions and nucleons.  
The implications of the Lagrangian \eqref{eq:q_Lcpv} (specialized to $n_f=2$ by neglecting the strange quark)  on the interactions between nucleons and pions can be studied using chiral perturbation theory ($\chi$PT).
The kinetic part of the QCD Lagrangian is invariant under the global chiral group $SU(2)_L \times SU(2)_R$ which is spontaneously broken to the isospin subgroup $SU(2)_V$ in the ground state. This leads to the emergence of the triplet of pseudo-Nambu-Goldstone bosons, the pion triplet, 
whose interactions are dictated by chiral symmetry. The pion mass arises from the explicit breaking of chiral symmetry by the small quark masses (and, to lesser degree, charges). The $\bar\theta$ term and qCEDMs also break chiral symmetry and induce non-derivative CPV pion-nucleon interactions. 

Before continuing with the construction of the chiral Lagrangian, we note that the qCEDMs are related to CP-even qCMDMs by a $SU(2)_L \times SU(2)_R$ rotation. This implies that chiral symmetry relates the matrix elements of the isoscalar (isovector) qCEDM operator between $n_N$ nucleon and $n_\pi$ pions to those of the isovector (isoscalar) qCMDM operator with $n_N$ nucleons and $n_\pi \mp 1$ pions. In particular, CPV pion-nucleon couplings induced by the qCEDMs
can be expressed in terms of corrections to the nucleon and pion masses and mass splittings induced by the qCMDMs.
This is analogous to the relation between the pion-nucleon coupling $\bar g_0$ induced by the QCD $\bar\theta$ term and the component of the nucleon mass splitting induced by the quark masses \cite{Crewther:1979pi,Mereghetti:2010tp}.

These relations between CP-even mass corrections and CPV pion-nucleon couplings can be straightforwardly derived using spurion techniques. The mass and dipole terms in Eq. \eqref{eq:q_Lcpv} break chiral symmetry, which can be formally restored by assigning the mass and dipole terms the following transformation properties under $SU(2)_L \times SU(2)_R$ rotations: 
\begin{eqnarray}\label{eq:2.1}
\mathcal M + i\left[ m_* \left(\bar\theta-\bar\theta_{\mathrm{ind}}\right) + r \tilde d_{CE}\right]  &\rightarrow& R \left\{ \mathcal M + i\left[ m_* \left(\bar\theta-\bar\theta_{\mathrm{ind}}\right) + r \tilde d_{CE}\right]\right\} L^{\dagger}\ , \nonumber \\
\tilde d_{CM}- i \tilde d_{CE} &\rightarrow&  R \left(\tilde d_{CM}- i \tilde d_{CE}\right) L^\dagger\ , 
\end{eqnarray}
where $L, R \in SU(2)_{L,R}$. The construction of the chiral Lagrangian now mimics that of the standard $\chi$PT Lagrangian and we refer to Refs. \cite{deVries:2012ab,Bsaisou:2014oka,Mereghetti:2010tp,Gasser:1984gg,Borasoy:2000pq, deVries:2015gea} for more details. We introduce
\begin{eqnarray}\label{U}
U(\pi) = u(\pi)^2 = \exp\left(  \frac{ i \boldpi \cdot \boldtau}{F_0}\right),
\end{eqnarray}
where $\boldpi$ denotes the pion triplet, $\boldtau$ the Pauli matrices, and $F_0$ is the pion decay constant in the chiral limit  (we use $F_\pi = 92.2$ MeV for  the physical decay constant),
and the spurion fields 
\begin{eqnarray}
\chi &=& 2 B \left( \mathcal M + i\left(m_* (\bar\theta - \bar\theta_{\textrm{ind}} ) + r \tilde{d}_{CE} \right) \right), \qquad  \tilde \chi = 2 \tilde B \left( \tilde{d}_{CM} - i \tilde{d}_{CE} \right), \nn \\
\chi_\pm & = &  u^{\dagger} \chi u^{\dagger} \pm u \chi^{\dagger} u\ , \qquad \tilde \chi_\pm  =   u^{\dagger} \tilde \chi u^{\dagger} \pm u \tilde \chi^{\dagger} u \ .\label{eq:2.2} 
\end{eqnarray}
The quantity $B$ is standardly related to the chiral condensate through the Gell-Mann--Oakes--Renner relation~\cite{GellMann:1968rz} while the new quantity, $\tilde{B}$ is related to the gluonic condensate:
\begin{align}
&B = -\frac{\langle 0 | \bar{q} q | 0 \rangle}{F_0^2}\, ,&
&\tilde{B} = -\frac{\langle 0 | \bar{q} g_s \sigma_{\mu\nu} G^{\mu\nu} q | 0 \rangle}{2 F_0^2}\, .&
\end{align}

The most important operators induced by the chiral-breaking mass and dipole terms are contributions to the pion and nucleon masses and mass splittings\footnote{We capitalize the low energy constants (LECs) $\tilde C_{1,5}$ to avoid confusion with the isoscalar and isovector CMDMs $\tilde c_{0,3}$.}
\begin{eqnarray}\label{eq:2.3}
\mathcal L_{} &=& \frac{F_0^2}{4}\left( \textrm{Tr}\, [U^{\dagger} \chi + U \chi^{\dagger}] +\textrm{Tr}\, [U^{\dagger} \tilde\chi + U \tilde\chi^{\dagger}]\right)
\nonumber\\
&&+ \left( c_1  \textrm{Tr}(\chi_+) + \tilde C_1  \textrm{Tr}(\tilde \chi_+) \right) \bar N N +
\bar N \left( c_5  \hat{\chi}_+  
+  \tilde C_5 
\hat{\tilde \chi}_+  \right)  N\ ,
\end{eqnarray}
where $N = (p\, n)^T$ is the nucleon doublet, and the hat denotes the traceless component of a chiral structure, e.g. $\hat \chi = \left(\chi  -\frac{1}{2} \textrm{Tr}(\chi)\right)$.
At this order, the $\chi$PT Lagrangian in presence of the $\bar\theta$ term and dipole operators is obtained from the standard $\chi$PT Lagrangian by changing $\chi$ to include the CPV terms and by 
noting that the CPV operators break chiral symmetry in the same pattern as the quark mass operator. This implies there will be a copy of each of the standard symmetry-breaking operators with $\chi \rightarrow \tilde \chi$ and a new, unknown accompanying LEC~\cite{deVries:2012ab,Bsaisou:2014oka}.
We have not included interactions that appear at the same order but play no role in our discussion.  By NDA, the LECs $c_{1,5}$ and $\tilde C_{1,5}$ scale as $\mathcal O(1/\Lambda_\chi)$.

The qCMDMs lead to a shift of the leading-order (LO) pion mass
\begin{equation}\label{eq:mpi}
\mpi^2 = m^2_{\pi,\mathrm{QCD}} + \tilde m_\pi^2 = 2 \left( B \bar m + \tilde B \tilde c_0\right)\, ,
\end{equation}
such that 
\begin{equation}\label{eq:r_mpisq}
r  = \frac{d m^2_\pi}{d \tilde c_0}\bigg/ \frac{d m^2_\pi}{d \bar m}= \frac{\tilde B}{B}  + \dots \ ,
\end{equation}
where the dots denote higher-order terms in the chiral expansion.
Furthermore, vacuum alignment ensures that at LO no purely mesonic interactions involving an odd number of pions survive \cite{deVries:2012ab}.
The nucleon mass terms are altered as well. We introduce the nucleon mass splitting, $\delta m_N = m_n -m_p$,  and the nucleon mass shift, $\Delta m_N = (m_n + m_p)/2- m^{0}_N$, where $m^0_N$ is the nucleon mass in the chiral limit. The following relations are obtained
\begin{eqnarray}
\delta m_N  &=& \delta m_{N,\mathrm{QCD}} + \tilde \delta  m_N = - 8 B (\bar m \varepsilon c_5 - \tilde c_3 r \tilde C_5)\ ,\\
 \Delta m_N &=& \Delta m_{N,\mathrm{QCD}} + \tilde \Delta  m_N = - 8 B (\bar m c_1 + \tilde c_0 r \tilde C_1)\, .
\end{eqnarray}
Finally, the presence of the CPV operators leads to various CPV pion-nucleon interactions
\begin{equation}
\mathcal L_\pi = - \frac{\bar g_0}{2 F_{\pi}} \bar N \boldtau\cdot \boldpi N  - \frac{\bar g_1}{2 F_{\pi}} \pi_0 \bar N  N  - \frac{\bar g_2}{2 F_{\pi}} \pi_0 \bar N  \tau^3 N + \ldots,
\end{equation}
where the dots indicate terms with additional pions that arise from expanding Eq.~\eqref{eq:2.3}. 
At tree-level the following relations immediately emerge
\begin{eqnarray}\label{tree}
\bar g_0 &=& \left[ \frac{\tilde \delta m_N}{\tilde c_3}+ \left(\frac{\tilde m^2_\pi}{\tilde c_0}\right)\left( \frac{\bar m }{m^2_{\pi,\mathrm{QCD}}}\right)\left(\frac{\delta m_{N,\mathrm{QCD}}}{\bar m \varepsilon}\right)\right] \tilde d_0 + \delta m_{N,\mathrm{QCD}} \frac{1-\varepsilon^2}{2\varepsilon} \left(\bar\theta-\bar\theta_{\mathrm{ind}}\right)\ ,\nonumber \\
\bar g_1 &=& -2 \left[ \frac{\tilde \Delta m_N}{\tilde c_0} - \left(\frac{\tilde m^2_\pi}{\tilde c_0}\right)\left( \frac{\bar m }{m^2_{\pi,\mathrm{QCD}}}\right)\left(\frac{\Delta m_{N,\mathrm{QCD}}}{\bar m} \right)\right]\tilde d_3 \ ,
\end{eqnarray}
while $\bar g_2=0$ at this order. These LO relations have been identified in Refs.~\cite{Pospelov:2001ys,deVries:2012ab}. 
It is useful to look at the relations in a bit more detail. The right-hand side depends on QCD quantities such as the pion and nucleon mass induced by the average quark mass, and the nucleon mass splitting induced by the quark mass difference. These quantities are known nowadays to rather good precision\footnote{It must be said that a discrepancy has emerged in the precise value of the nucleon sigma term. Recent LQCD calculations \cite{Durr:2015dna} seem to provide values at odd with pion scattering lengths \cite{Hoferichter:2016ocj}.}. In addition, the right-hand side depends on three new quantities given by pion and nucleon mass shifts from the isoscalar qCMDM, $\tilde m_\pi^2$ and $\tilde \Delta m_N$, and the nucleon mass splitting from the isovector qCMDM, $\tilde \delta m_N$. The practical advantage of Eq.~\eqref{tree} is that the latter are spectroscopic quantities that can be calculated readily with lattice QCD, while direct calculations of the CPV three-point functions are more difficult.

\begin{figure}
\center
\includegraphics[width=14cm]{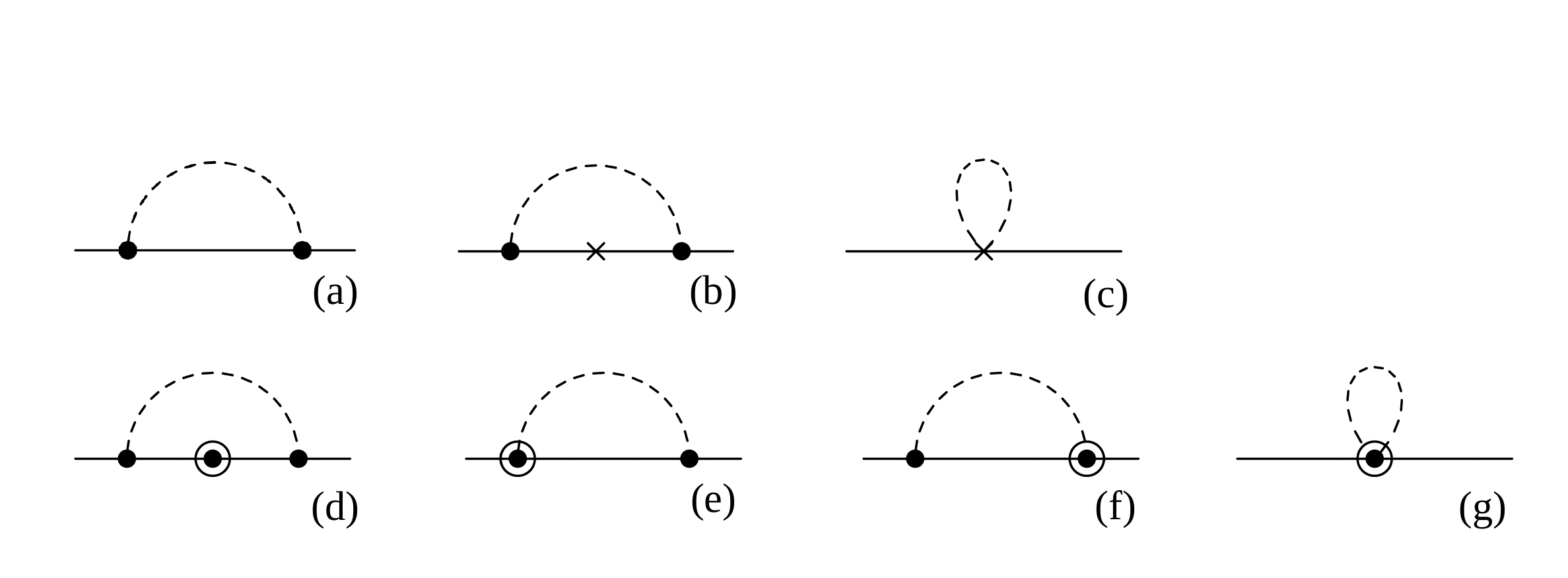}
\caption{NLO correction (diagram $(a)$) and N${}^2$LO corrections (diagrams $(b)$--$(g)$) to the nucleon mass
and mass splitting.
Plain and dashed lines denote nucleons and pions. The dots denote vertices in the leading chiral Lagrangian. 
Circled dotted vertices denote $SU(2)$ invariant couplings in the $\mathcal O(p^2)$ baryon chiral Lagrangian. Crosses denote insertions of the chiral-symmetry breaking nucleon interactions in Eq. \eqref{eq:2.3}.}\label{massNLO}
\end{figure}

At first sight, Eq.~\eqref{tree} provides a promising way to calculate the CPV pion-nucleon couplings. However, the relations are based on LO $\chi$PT and higher-order corrections could spoil them. In fact, Ref.~\cite{Seng:2016pfd} demonstrated that the $\bar g_1$ relation is broken at next-to-leading order (NLO) and obtains $\mathcal O(100\%)$ corrections. The violation can be easily seen by considering NLO corrections to the left- and right-hand sides in Eq.~\eqref{tree}. The first corrections to $\bar g_{0,1,2}$ only appear at N${}^2$LO \cite{deVries:2012ab}, while the nucleon mass obtains an NLO correction from Fig. \ref{massNLO}(a)
\begin{equation}\label{eq:mn_mpi3}
\Delta m_N^{(1)} = -\frac{3\pi g_A^2}{2} \frac{m_\pi^3}{(4 \pi F_0)^2} \ ,
\end{equation} 
such that the $\bar g_1$ relation is explicitly violated and that the practical use of Eq.~\eqref{tree} in obtaining precise values of the CPV coupling constants is limited\footnote{Ref.~\cite{Seng:2016pfd} found that the $\bar g_0$ relation was only affected at N${}^2$LO and numerically more stable.}. 
This non-analytic quark mass correction, Eq.~\eqref{eq:mn_mpi3}, also seems to spoil the convergence of $SU(2)$ chiral perturbation theory for the nucleon mass when compared to LQCD calculations at unphysically heavy quark masses~\cite{WalkerLoud:2008bp,WalkerLoud:2008pj,Walker-Loud:2013yua,Walker-Loud:2014iea}.

We point out here that it is possible to modify the tree-level relations in Eq.~\eqref{tree} such that they not only survive all NLO corrections but also the dominant part of N${}^2$LO corrections. The relations then obtain similar precision as those found for the QCD $\tb$ term, $\mathcal O(15\%)$. The modified relations are 
\begin{eqnarray}\label{eq:g0g1_mod}
\bar g_0 &=& \tilde d_0\left( \frac{d }{d \tilde c_3} + r \frac{d}{d (\bar m \varepsilon)}\right)\delta m_N  + \delta m_{N,\textrm{QCD}} \frac{1-\varepsilon^2}{2\varepsilon} \left(\bar\theta-\bar\theta_{\mathrm{ind}}\right)\ ,\nonumber \\
\bar g_1 &=& -2 \tilde d_3 \left( \frac{d}{d \tilde c_0} - r \frac{d}{d \bar m} \right)\Delta m_N\ ,
\end{eqnarray}
with $r$ given in Eq.~\eqref{eq:r_mpisq}.
At LO in the chiral expansion the modified relations are identical to those in Eq.~\eqref{tree}.  
We can immediately check that the new relations are preserved at NLO. The NLO correction to the right-hand side of the $\bar g_1$ relation is a function of the pion mass only and it is easy to see that
\begin{equation}\label{trick}
\left( \frac{d}{d \tilde c_0} - r \frac{d}{d \bar m} \right) f(\mpi^2) = \left( \frac{d\mpi^2}{d \tilde c_0}\frac{d}{d \mpi^2} - \frac{d\mpi^2}{d \tilde c_0}\frac{d}{d \mpi^2 }\right) f(\mpi^2) =0\, .
\end{equation}

\begin{figure}
\center
\includegraphics[width=15cm]{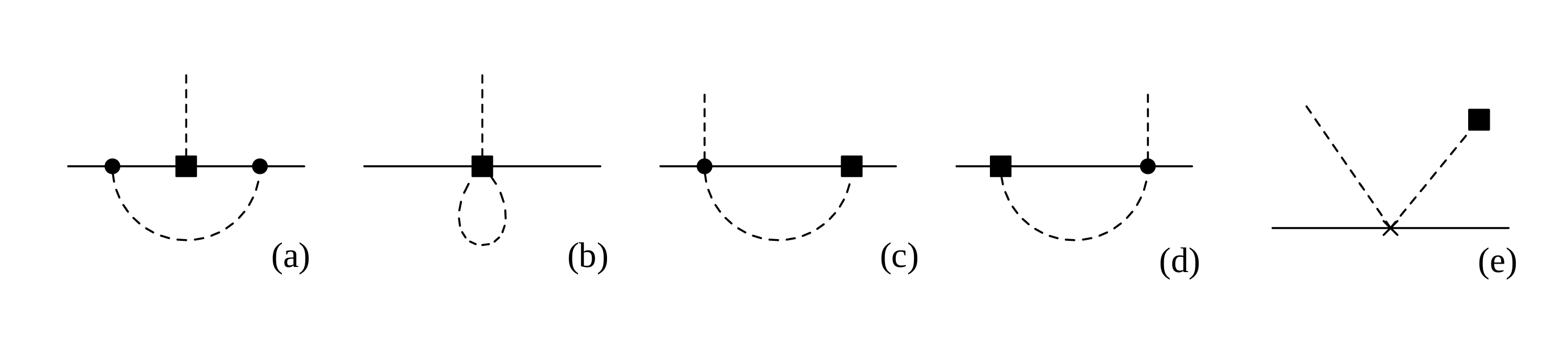}
\caption{N${}^2$LO contributions to $\bar g_0$ and $\bar g_1$. Squares denote CPV pion-nucleon couplings. Other notation is as in Fig. \ref{massNLO}}\label{FFn2lo}
\end{figure}

We now turn to N${}^2$LO corrections which are potentially dangerous as they can be enhanced by large logarithms \cite{Seng:2016pfd}. We focus on the PQ scenario to reduce the number of terms, but our conclusions do not depend on this \cite{deVries:2015una}. The terms in standard $\chi$PT can be found in Ref.~\cite{Gasser:1983yg,Fettes:2000gb} and here we list the relevant terms at $\mathcal O(p^4)$ (the QCD LECs $l_{3,4}$ are defined to be precisely those of the original work~\cite{Gasser:1983yg})
\begin{eqnarray}
\mathcal L &=& 
	\frac{l_3 +l_4}{16} \textrm{Tr}(\chi_+)^2
	+\frac{l_4}{8} \textrm{Tr}(\chi_+) \textrm{Tr} (\partial_\mu U \partial^\mu U^\dagger)
	- \frac{l_7}{16} \textrm{Tr}( \chi_- )^2 
	+ \frac{2\tilde{l}_3+\tilde{l}_4+l_4}{16} \textrm{Tr} ({\chi}_+) \, \textrm{Tr} (\tilde{\chi}_+)   
\nn  
\\ & &
	+\frac{\tilde{l}_4}{8} \textrm{Tr}(\tilde{\chi}_+) \textrm{Tr} (\partial_\mu U \partial^\mu U^\dagger)
	- \frac{\tilde{l}_7}{8} \textrm{Tr} ( \chi_-) \textrm{Tr} (\tilde\chi_-) 
\nn
\\ & & + \bar N \left[ e_1 \textrm{Tr} ( \chi_+  )^2  
	+ e_2  \hat{\chi}_+ \, \textrm{Tr} ( \chi_+ )  
	+ e_3 \textrm{Tr} ( \hat{\chi}_{+} \hat{\chi}_{+}) 
	+ e_4\,  \hat{\chi}_- \, \textrm{Tr}( \chi_- ) 
	+ \frac{e_5}{4} \textrm{Tr} ( \chi_+^2 - \chi_-^2)  \right. \nn 
\\ & & \qquad \left. 
	- \frac{e_6}{4} \left( \textrm{Tr}(\chi_-^2) 
	- \textrm{Tr}(\chi_-)^2 
	+ \textrm{Tr}(\chi^2_+) 
	- \textrm{Tr}(\chi_+)^2   \right)
	\right] N \nn 
\\ & & 
	+ \bar N \left[ 2\tilde{e}_1 \textrm{Tr} ( \chi_+  )\, \textrm{Tr} ( \tilde{\chi}_+  )   
	+ \tilde{e}_{2a} \, \hat{\tilde{\chi}}_+ \, \textrm{Tr} ( \chi_+ ) 
	+ \tilde{e}_{2b}\hat{{\chi}}_+ \, \textrm{Tr} ( \tilde\chi_+ )  
	+ 2 \tilde{e}_3 \textrm{Tr} ( \hat{\chi}_{+} \hat{\tilde{\chi}}_{+})  \nn \right. 
\\ & & \qquad \left. 
	+ \tilde{e}_{4a}\,   \hat{\tilde{\chi}}_- \, \textrm{Tr}( \chi_- )  
	+ \tilde{e}_{4b}\hat{\chi}_- \, \textrm{Tr}( \tilde{\chi}_- ) 
	+ \frac{\tilde{e}_5}{2} \textrm{Tr} ( \chi_+ \tilde{\chi}_+ 
	- \chi_- \tilde{\chi}_-)  \right. \nn 
\\ & & \qquad \left. 
	- \frac{\tilde{e}_6}{2} \left( \textrm{Tr}(\chi_- \tilde{\chi}_-) 
	- \textrm{Tr}(\chi_-) \textrm{Tr}(\tilde{\chi}_-) 
	+ \textrm{Tr}(\chi_+\, \tilde{\chi}_+) 
	- \textrm{Tr}(\chi_+) \textrm{Tr}(\tilde{\chi}_+)    \right)
	\right] N\ . 
\label{eq:subleading}
\end{eqnarray}
The $l_i,\tilde{l}_i$ ($i=3,4$) terms correct the pion wavefunction renormalization, the pion mass, and $F_\pi$. Combining them with the one-loop corrections gives (neglecting terms quadratic in isospin breaking):
\begin{eqnarray}
Z_\pi&=&1-\frac{m_\pi^2}{24\pi^2F_0^2}L_\pi-\frac{4B\bar{m}}{F_0^2}l_4^r-\frac{4\tilde{B}\tilde{c}_0}{F_0^2}\tilde{l}_4^r+\frac{m_\pi^2}{12\pi^2F_0^2}L_\epsilon\ , \nonumber\\
m_\pi^2&=&(m_\pi^2)_{\mathrm{tree}}\left\{1-\frac{m_\pi^2}{32\pi^2F_0^2}L_\pi\right\}+\frac{8B^2\bar{m}^2}{F_0^2}l_3^r+\frac{16B\tilde{B}\bar{m}\tilde{c}_0}{F_0^2}\tilde{l}_3^r\ , \nonumber\\
F_\pi&=&F_0\left\{1+\frac{m_\pi^2}{16\pi^2F_0^2}L_\pi+\frac{2B\bar{m}}{F_0^2}l_4^r+\frac{2\tilde{B}\tilde{c}_0}{F_0^2}\tilde{l}_4^r\right\}\ ,
\end{eqnarray}
where $L_\pi = \log \frac{\mu^2}{\mpi^2}$. We applied dimensional regularization in $d = 4-2\epsilon$ dimensions and absorbed $L_\epsilon =  1/\epsilon - \gamma_E +\log 4\pi +1$ 
in the renormalization-scale ($\mu$) dependent counterterms
\begin{eqnarray}
&&l_3^r=l_3-\frac{1}{64\pi^2}L_\epsilon\ ,\qquad\tilde{l}_3^r=\tilde{l}_3-\frac{1}{64\pi^2}L_\epsilon\ ,\nonumber\\
&&l_4^r=l_4+\frac{1}{16\pi^2}L_\epsilon\ ,\qquad \tilde{l}_4^r=\tilde{l}_4+\frac{1}{16\pi^2}L_\epsilon\ ,
\end{eqnarray}	
in order to make $m_\pi^2$ and $F_\pi$ finite and scale-independent. The wavefunction renormalization $Z_\pi$, on the other hand, remains infinite and scale-dependent.
$l_7$ and $\tilde l_7$ provide analogous corrections to the pion mass splitting. In addition, these operators induce a pion tadpole, which we treat in perturbation theory (see Fig.~\ref{FFn2lo}(e)).
The $\tilde e_i$ terms in Eq. \eqref{eq:subleading} induce corrections to the nucleon mass terms and CPV pion-nucleon couplings,
suppressed by $m^2_\pi/\Lambda_\chi^2$ with respect to LO terms.

We now investigate the N${}^2$LO corrections to $\Delta m_N$ and $\delta m_N$ depicted in Fig.~\ref{massNLO}. Diagrams $(d)$--$(g)$ contribute only to $\Delta m_N$ and contain NLO chiral-invariant vertices, including recoil corrections
to the nucleon propagator and the axial pion-nucleon couplings $g_A$, and chiral-invariant nucleon-two-pion couplings. As was the case for the NLO correction in Fig.~\ref{massNLO}, these diagrams only depend on $\bar m$ and $\tilde c_0$ through the pion mass, and Eq.~\eqref{trick} proves that they do not contribute to the right-hand side of Eq.~\eqref{eq:g0g1_mod}.
The only relevant corrections to the nucleon masses then originate in diagrams $(b)$ and $(c)$ \cite{WalkerLoud:2009nf}%
\begin{eqnarray}\label{N2LOmN}
\delta m_N &=& 
	(\delta m_N)_{\mathrm{tree}} f_0(m^2_\pi)  + \delta m_N^{\textrm{ct}} = 
 	(\delta m_N)_{\mathrm{tree}}\left\{1 
		+ \frac{\mpi^2}{2 (4\pi F_0)^2}\left[ \left(1 +6 g_A^2\right) L_\pi  
			- 4 g_A^2\right]
		\right\}  \nn 
\\ & & 
	- 64 B^2 \left( \bar m^2 \varepsilon \, e^r_{2} 
		-r \bar m \tilde c_3 \, \tilde{e}^r_{2a} 
		+r \bar m \varepsilon \tilde c_0\,\tilde e^r_{2b}  \right) \, ,
\nonumber\\
\Delta m_N &=& 
	(\Delta m_N)_{\mathrm{tree}} f_1(m^2_\pi) + \Delta m_N^{\textrm{ct}}  = 
	(\Delta m_N)_{\mathrm{tree}}\left[1 
		+ \frac{3\mpi^2}{2 (4\pi F_0)^2} L_\pi \right] \nn \\ 
& & 
	- 32 B^2 \left( \bar m^2 \left(2 e^r_1 + \frac{e_5}{4}+ \frac{e_6}{4} \right) 
		+ 2 r \bar m \tilde c_0 \left(2  \tilde e^r_1 + \frac{\tilde e_5}{4}+\frac{\tilde e_6}{4} \right)\right) \, .
\end{eqnarray}
The renormalized counterterms are
\begin{eqnarray}\label{LECren}
e_1^r &=& e_1 + \frac{3 c_1}{8 (4\pi F_0)^2} L_\epsilon\ , \qquad 
\tilde{e}_1^r = \tilde{e}_1 + \frac{3 (c_1 + \tilde C_1) }{16 (4\pi F_0)^2} L_\epsilon \ ,
\nn \\
e_2^r  &=& e_2 + \frac{(1 + 6 g_A^2) c_5}{ 8 (4\pi F_0)^2} L_\epsilon\ , \quad 
\tilde{e}_{2a}^r =\tilde{e}_{2a} + \frac{(1 + 6 g_A^2) \tilde{C}_5}{ 8 (4\pi F_0)^2}  L_\epsilon\ , \quad 
\tilde{e}_{2b}^r =\tilde{e}_{2b} + \frac{(1 + 6 g_A^2) c_5}{ 8 (4\pi F_0)^2} L_\epsilon\ . 
\end{eqnarray}
The LECs $e_3$, $e_5$, and $e_6$, and the analogous ones induced by $\tilde c_0$, are not renormalized.
One way to see this is that these operators induce corrections to $\Delta m_N$ that are quadratic in isospin breaking, for which there is no loop divergence at this order.
$e_4$, $\tilde e_{4a}$, and $\tilde e_{4b}$ do not affect the nucleon mass.

The N${}^2$LO contributions to $\bar g_{0,1,2}$ are given by
\cite{deVries:2012ab} 
\begin{eqnarray}\label{eq:N2LOg}
\bar g_0 &=& 
	(\bar g_0)_{\mathrm{tree}} \left\{1 
		+ \frac{\mpi^2}{2 (4\pi F_0)^2}\left[ \left(1 +6 g_A^2\right) L_\pi 
		- 4 g_A^2\right]\right\}
	+64 B^2  r \bar m \tilde{d}_0  \left( \tilde{e}^r_{2a} -  e^r_2 + \tilde e_{4b} - e_4 \right)  
\nn \\
\bar g_1 &=& (\bar g_1)_{\mathrm{tree}} \left[1 
	+ \frac{3\mpi^2}{2 (4\pi F_0)^2} L_\pi  \right] 
	+ 128 B^2 r \left( 2 \bar m\, \tilde d_3   (\tilde e^r_1 - e^r_1) 
	- \bar m \varepsilon \, \tilde d_0 \,(\tilde e_3 - e_3)  \right)
\nn \\ 
\bar g_2 &=& 
	64 B^2 r\, \bar m \varepsilon  \tilde d_3\, \left(e_2 - \tilde{e}_{2b} + e_4 - \tilde{e}_{4a} \right)
\end{eqnarray}
Compared to Ref. \cite{deVries:2012ab}, we have included the nucleon and pion wavefunction renormalization, and the one-loop corrections to $F_0$. 
The loop functions appearing in Eqs. \eqref{N2LOmN} and \eqref{eq:N2LOg} are exactly the same, and once the divergences in $m_N$ and $\delta m_N$ are subtracted as in Eq. \eqref{LECren}, the pion-nucleon couplings are 
also renormalized. Thus $e_4- \tilde e_{4 a}$ and $e_4- \tilde e_{4 b}$ are finite and $\mu$-independent.
At this order, the first contribution to  $\bar g_2$ appears which depends purely on counterterms in Eq. \eqref{eq:subleading}. Eq. \eqref{LECren} guarantees  that the combination $e_2 - \tilde e_{2b}$ is finite and $\mu$-independent.

We now demonstrate that the N${}^2$LO loop corrections and the counterterms $e_1$, $\tilde e_1$, $e_2$, $\tilde e_{2a}$  preserve the relations in Eq.~\eqref{eq:g0g1_mod}. We start with the $\bar g_0$ relation. Because the loop function $f_0(m^2_\pi)$ does not depend on $\bar m \varepsilon$, the derivatives $(d/d c_3 + rd/d (\bar m \varepsilon))$ only act on $(\delta m_N)_{\mathrm{tree}}$. The same function appears in both the correction to $\bar g_0$ and $\delta m_N$ such that Eq.~\eqref{eq:g0g1_mod} is preserved. 
For almost identical reasons the $\bar g_1$ relation is preserved by loop corrections once Eq.~\eqref{trick} is used to eliminate the derivatives acting on $f_1 (\mpi^2)$. An explicit check shows that the contributions of $e_1$, $\tilde e_1$, $e_2$ and $\tilde e_{2a}$ satisfy Eq. \eqref{eq:g0g1_mod}.

The remaining counterterms, $e_{3-6}$ lead to the following violations of the relations
\begin{eqnarray}\label{violations1}
\delta \bar g_0^{\textrm{ct}} &=& 64 B^2 \, \bar m  r \, \tilde d_0 \left( \tilde{e}_{4b} - e_4\right) \\
\delta \bar g_1^{\textrm{ct}} &=& 128 B^2 r \left( \bar m \, \tilde d_3 \left( \frac{e_5}{4} -\frac{\tilde e_5}{4} +  \frac{e_6}{4} - \frac{\tilde e_6}{4} \right)  +  \bar m \varepsilon\, \tilde d_0 (e_3 - \tilde e_3)\right).
\end{eqnarray}
At the same time, we find a N${}^2$LO correction to $r$ 
\begin{equation}
r = r^{(0)} \left( 1 + \frac{4 m^2_\pi}{F_0^2} \left( \tilde{l}_3  - l_3\right)  \right),
\end{equation}
and corrections to $\bar g_{0,1}$ from the tadpole diagrams in Fig.~\ref{FFn2lo}. These can be combined into 
\begin{eqnarray}\label{violations2}
\delta \bar g_0^{\textrm{tad}} &=&  - r  \frac{4 m^2_\pi}{F_0^2} \frac{d \delta m_N}{d \bar m \varepsilon} \left( 
\tilde d_0 (\tilde l_3 - l_3) + \frac{\varepsilon}{2} \tilde d_3 \left(\tilde{l}_3 - l_3 +\frac{\tilde{l}_4}{2}-\frac{l_4}{2}\right) + \frac{\varepsilon^2}{2} \tilde d_0 \left( \tilde{l}_7 - l_7\right) \right)\ , \nonumber\\
\delta \bar g_1^{\textrm{tad}} &=&  - r \frac{4 m^2_\pi}{F_0^2} \frac{d \Delta m_N}{d \bar m} \left(  \tilde d_3 \left(\tilde{l}_3 - l_3-\frac{\tilde{l}_4}{2}+\frac{l_4}{2}\right) -  \varepsilon \tilde d_0 \left( \tilde{l}_7 - l_7\right) \right) \ .
\end{eqnarray}
From the infinity-subtraction of the counterterms one finds that $\delta \tilde{g}_{0,1}^{\mathrm{tad}}$ are finite and scale-independent. We cannot determine the size of the violations exactly because of the appearance of unknown counterterms, but  we can get a reasonable estimate by looking at a piece we do control. 
For instance, if we neglect isospin breaking, $\varepsilon = 0$, and the derivatives with respect to $\tilde c_{0,3}$ in Eq. \eqref{eq:g0g1_mod}, the ratio of the tadpole and tree level couplings is
\begin{eqnarray}
\frac{\delta \bar g_0^{\textrm{tad}}}{\bar g_0} &=&  \frac{m^2_\pi}{(4\pi F_0)^2} (\bar{\tilde l}_3 - \bar l_3), \qquad 
\frac{\delta \bar g_1^{\textrm{tad}}}{\bar g_1} =  \frac{m^2_\pi}{(4\pi F_0)^2} \left( \frac{1}{2} (\bar{\tilde l}_3 - \bar l_3)  + (\bar{\tilde l}_4 - \bar l_4) \right), 
\end{eqnarray}
where we introduced the scale independent counterterms $\bar l_{3,4}$  \cite{Gasser:1983yg}
\begin{align}
&l_3^r(\mu) = \frac{-1/2}{32\pi^2} \left[ \bar{l}_3 + \ln \left(\frac{m_\pi^2}{\mu^2}\right)\right]\, ,&
&l_4^r(\mu) = \frac{2}{32\pi^2} \left[ \bar{l}_4 + \ln \left(\frac{m_\pi^2}{\mu^2}\right)\right]\, ,&
\end{align}
and their analogs $\bar{\tilde l}_{3,4}$ for the qCMDM-induced operators. 
The FLAG review~\cite{Aoki:2016frl} provides the estimates 
\begin{align}
&\bar{l}_3 = 2.81(64),\, &
&\bar{l}_4 = 4.10(45),\, &
\end{align}
from which, barring large cancellations between $\bar l_{3,4}$ and $\bar{\tilde l}_{3,4}$, we obtain
\begin{align}
&\frac{\delta \bar g_0}{\bar g^{}_0} \simeq -0.04,\, &
&\frac{\delta \bar g_1}{\bar g^{}_1} \simeq -0.08.\, &
\end{align}
The other corrections in Eq.~\eqref{violations2} are expected to be even smaller because of additional factors of $\varepsilon \simeq 0.3$. Clearly, this is not a full determination of the theoretical uncertainty of the relations in Eq.~\eqref{eq:g0g1_mod}, but there is no reason to expect the remaining violations in Eqs.~\eqref{violations1} and \eqref{violations2}  to be significantly larger either. To be on the conservative side, we follow Ref.~\cite{deVries:2015una} and assign a $20\%$ intrinsic uncertainty to the relations. Considering that it is unlikely that lattice determinations, at least in the near future,  of the right-hand side of Eq.~\eqref{eq:g0g1_mod} will reach this accuracy, for practical purposes the relations can be treated as exact.

%
\subsection{$SU(3)$ flavor-breaking corrections}
We now briefly extend the discussion to $SU(3)$ $\chi$PT to investigate the role of the strange CEDM.
Most of the formalism follows directly
from  the previous section, with the inclusion of  strange hadrons. For example, the LO Lagrangian induced by mass terms and dipole operators is 
\begin{eqnarray}\label{eq:3.1}
\mathcal L_{} &=& \frac{F_0^2}{4}\left( \textrm{Tr}\, [U^{\dagger} \chi + U \chi^{\dagger}] +\textrm{Tr}\, [U^{\dagger} \tilde\chi + U \tilde\chi^{\dagger}]\right) + b_0 \textrm{Tr} \left(\bar B B\right) \textrm{Tr} \chi_+  +  b_D \textrm{Tr}\left(\bar B \{ \chi_+, B \}  \right) + b_F \textrm{Tr}\left(\bar B [ \chi_+, B ]  \right) \nonumber \\
 &  & +  \tilde b_0 \textrm{Tr} \left(\bar B B\right) \textrm{Tr} \tilde \chi_+  +  \tilde b_D \textrm{Tr}\left(\bar B \{ \tilde \chi_+, B \}  \right) + \tilde b_F \textrm{Tr}\left(\bar B [\tilde  \chi_+, B ]  \right)\, ,
\end{eqnarray}
where we adopted the notation of Ref.~\cite{deVries:2015una} for the meson ($U$) and baryon ($B$) octet fields.

The interactions in Eq. \eqref{eq:3.1} affect  meson masses and mixings, and induce 
both baryon masses and splittings \cite{Frink:2004ic} and CPV meson-nucleon vertices. 
Using Eq. \eqref{eq:3.1}, it is again possible to express $\bar g_{0,1}$ in terms of the nucleon masses. 
Neglecting terms proportional to the $\eta-\pi$ mixing angle $\phi$, which are formally LO in the $SU(3)$ expansion but, being $\phi \sim \varepsilon \bar m/\bar m_s \sim \mathcal O(10^{-2})$, are numerically very small,
we can generalized Eq. \eqref{eq:g0g1_mod} to
\begin{eqnarray}\label{modStrange2}
\bar g_0 &=& \tilde d_0\left( \frac{d }{d \tilde c_3} + r \frac{d}{d (\bar m \varepsilon)}\right)\delta m_N + 
 \delta m_{N,\mathrm{QCD}} \frac{1-\varepsilon^2}{2\varepsilon} \left(\bar\theta-\bar\theta_{\mathrm{ind}}\right)\ ,\nonumber \\
\bar g_1 &=& -2 \tilde d_3 \left( \frac{d}{d \tilde c_0} - r \frac{d}{d \bar m} \right)\Delta m_N+ 4 \frac{\phi}{\sqrt{3}}\left[ \tilde d_s \left( \frac{d}{d \tilde c_s} - r \frac{d}{d m_s} \right)\right]\Delta m_N \ ,
\end{eqnarray}
where $\phi/\sqrt{3} = \bar m \varepsilon/(2 (m_s - \bar m))$.
The expression for $\bar g_0$ is the same as in Eq. \eqref{eq:g0g1_mod}. In the case of $\bar g_1$, we obtain an additional contribution, proportional to  $\phi \times \tilde d_s$. 
Since in most BSM scenarios the qCEDMs scale with the quark mass,  $\phi\times\tilde d_s$ does not vanish in the large $m_s$ limit, and it is not suppressed with respect to $\tilde d_{0,3}$.
$\bar g_1$ can therefore obtain significant contributions from the strange CEDM \cite{Hisano:2004tf}, which is particularly interesting because $\bar g_1$ provides a major contribution to the EDMs of nuclei and diamagnetic atoms.
In order to quantify the size of $\bar g_1$, we require, apart from the matrix elements discussed in the $SU(2)$ case, the strange content of the nucleon $m_s (d\Delta m_N/d m_s)$, known to $25\%$ accuracy \cite{Junnarkar:2013ac}, and the analogous dependence of the nucleon mass on the strange CMDM $ (d\Delta m_N/d \tilde c_s)$.

We now discuss higher-order corrections to Eq. \eqref{modStrange2}.
The case of the QCD $\bar\theta$ term was discussed in detail in Ref. \cite{deVries:2015una}, and all the derived conclusions apply here with the modification $\bar\theta \rightarrow \bar\theta - \bar\theta_{\textrm{ind}}$. In particular, for the $\bar\theta$ term the relation connecting $\bar g_0$ to $\delta m_N$ is protected from higher-order corrections. 
As far as the remaining terms in Eq. \eqref{modStrange2} are concerned, the derivative nature of the relations ensures that all NLO corrections are safe, as in $SU(2)$ $
\chi$PT. 
Similarly, the bulk of N${}^2$LO corrections affect the left- and right-hand sides of Eq.~\eqref{modStrange2} in the same way. Some violations do appear in the form of N${}^2$LO counterterms, higher-order isospin-violating corrections, and tadpole contributions. An estimate of these corrections following the strategy of Ref.~\cite{deVries:2015una} indicates that $SU(3)$ corrections to $\bar g_{0,1}$ induced by  $\tilde d_{0,3}$ relations are not larger than the $SU(2)$ violations discussed above.  
The dependence of $\bar g_1$ on $\tilde d_s$ is more uncertain. Loop corrections respect Eq. \eqref{modStrange2}, however in this case, since the LO contribution is suppressed by $\phi$, the N${}^2$LO counterterms 
are suppressed by only $m^2_K/\Lambda^2_\chi$ instead of $m_\pi^2/\Lambda^2_\chi$. The N${}^2$LO counterterms could therefore induce significant corrections, up to $\sim 50\%$. 
Even though the relation between $\bar g_1$ and $\tilde d_s$  in Eq.~\eqref{modStrange2} can not be put on the same solid theoretical footing as Eq.~\eqref{eq:g0g1_mod},
it still provides an interesting way to estimate the dependence of $\bar g_1$ on $\tilde d_s$.

%
\subsection{Hadron spectroscopy in the presence of higher dimension operators from Lattice QCD}
In order to utilize the relations Eq.~\eqref{eq:g0g1_mod}, one must perform spectroscopic LQCD calculations of the nucleon masses in the presence of higher-dimensional operators, as well as calculations of the standard QCD sigma-terms, 
\begin{equation}\label{eq:fh_matrix_elements}
\begin{array}{cc}
\textrm{QCD sigma terms}& \textrm{qCMDM sigma terms}
\\
\bar{m}\langle p | \bar{q} q | p \rangle =
	\bar{m} \frac{d}{d \bar{m}} \Delta m_N \Big|_{\bar{m}}&
\frac{\tilde{d}_3}{2} \langle p | g_s \bar{q} \sigma_{\mu\nu} G^{\mu\nu} q | p \rangle  =
	\tilde{d}_3 \left(\frac{d}{d \tilde{c}_0} \Delta m_N\right)\Big|_{\tilde{c}_0=0}
\\
\bar{m} \varepsilon \langle p | \bar{q} \tau_3 q | p \rangle =
	\bar{m}\varepsilon \frac{d}{d(\bar{m}\varepsilon)} \delta m_N \Big|_{\bar{m}\varepsilon}&
\frac{\tilde{d}_0}{2} \langle p | g_s \bar{q}  \sigma_{\mu\nu} G^{\mu\nu} \tau_3 q | p \rangle =
	\tilde{d}_0 \left(\frac{d}{d \tilde{c}_3} \delta m_N\right) \Big|_{\tilde{c}_3=0}\ ,
\end{array}\, 
\end{equation}
and of the ratio of vacuum matrix elements, $r$, defined in Eq. \eqref{eq:r_mpisq}.
The determination of the QCD sigma-term operators are now staple lattice QCD calculations, see Ref.~\cite{collins:latt2016} for the most recent review.
Matrix elements of these new operators can be determined through the linear response of the theory to external sources coupled to these operators.
An efficient means of performing these matrix element calculations is described in Ref.~\cite{Bouchard:2016heu}.
The LQCD calculations are in principle, straightforward, with a few technical details which must be addressed.

The chromo-magnetic operators, $\bar q \sigma^{\mu\nu} G_{\mu\nu} \tilde{d}_{CM} q$, break chiral symmetry in the same manner as the quark mass operators, and therefore these operators will mix under renormalization. In the $\overline{\textrm{MS}}$ scheme, the mixing is proportional to $m_q^2$, and thus small.  
However, if a lattice regularization scheme is used that breaks chiral symmetry, there is another source of chiral symmetry breaking which persists at finite lattice spacing and can be large.
Near the continuum limit, the effects of this chiral symmetry breaking due to the lattice discretization are captured at LO in the Symanzik expansion~\cite{Symanzik:1983dc} through a dimension-five \textit{clover} operator, which is in fact the chromo-magnetic operator, $a \bar q \sigma^{\mu\nu} G_{\mu\nu} q$~\cite{Sheikholeslami:1985ij}, where $a\sim \Lambda^{-1}$ is the discretization scale.
Often, this clover operator is added to the numerical calculation with a coefficient $c_{SW}$  that is tuned to remove the leading $\mathcal{O}(a)$ discretization effects.
In such a lattice calculation, the addition of the qCMDM operator would amount to adding a correction to the tuned value of the coefficient $c_{SW} \rightarrow c_{SW} + \delta c_{SW}$,
\begin{equation}
\mathcal{L} \supset a (c_{SW}+\delta c_{SW}) \bar{q} \sigma^{\mu\nu} G_{\mu\nu} q
	= a \left(c_{SW}+\frac{\tilde{d}_3g_s}{2a}\right) \bar{q} \sigma^{\mu\nu} G_{\mu\nu} q
\end{equation}
The challenge would be to tune the value of $\delta c_{SW} = g_s \tilde{d}_3/(2a)$ as one varies the gauge coupling (and hence the lattice spacing $a$) such that the corresponding value of the dimension-full coupling $\tilde{d}_3$ would remain fixed in physical units.
While possible in principle, this would involve a careful fine-tuning of the bare quark mass and the value of $\delta c_{SW}$ as all three operators would mix under renormalization.
This issue is restricted to the isoscalar qCMDM operator as the discretization effects are flavor blind.

Alternatively, one could utilize a lattice regularization scheme that respects chiral symmetry at finite lattice spacing~\cite{Ginsparg:1981bj,Luscher:1998pqa} such as Domain-Wall~\cite{Kaplan:1992bt,Shamir:1992im,Shamir:1993zy,Shamir:1998ww,Furman:1994ky} or overlap fermions~\cite{Narayanan:1992wx,Narayanan:1993sk,Narayanan:1993ss}.
With such a regularization scheme, there is still a complication of operator mixing, but the $c_{SW}$ operator is not present in the lattice action (or at least exponentially suppressed for Domain-Wall fermions).
The quark mass operators are lower-dimensional than the qCMDM operators so there will be a quadratic power-divergent mixing of $\bar{q}  q$ into $\bar{q} \sigma^{\mu\nu} G_{\mu\nu}  q$.
However, our proposed relations, Eq.~\eqref{eq:g0g1_mod} are constructed from the difference of two terms through the derivatives $\frac{d}{d\tilde c_0} - r \frac{d}{d \bar m}$, and it is straightforward to show that the quadratic power-divergent mixing cancels in this difference.  
If the lattice regularization respects chiral symmetry, there is no dimension-four operator that the qCMDM can mix into, 
and only logarithmic mixings need to be considered. It is easy to see that the logarithmic mixing with operators of the form $ \bar q m_q^2 q$ 
does not affect Eq. \eqref{eq:g0g1_mod}, and that, in the presence of chiral symmetry, the mixing of the qCMDM with $G_{\mu \nu} G^{\mu \nu}$
is proportional to the quark masses, and thus small. This leaves only the self-mixing of the qCMDM operator to be determined nonperturbatively. 
Provided the method described in Ref.~\cite{Bouchard:2016heu} is used in the LQCD calculations, these issues of operator mixing need only be dealt with when performing the renormalization, as the new qCMDM operators are not actually added in the lattice action, but are only used perturbatively to compute the matrix elements in Eq.~\eqref{eq:fh_matrix_elements}.

A final detail needs to be considered  when imposing renormalization conditions on the qCMDM and qCEDM operators. 
The CPV couplings $\bar g_0$ and $\bar g_1$ are independent of the renormalization scheme.
Eq.~\eqref{eq:g0g1_mod} guarantees this independence if the renormalization of the qCEDM and qCMDM operators is the same. 
This is true at one loop in the $\overline{\textrm{MS}}$ scheme,
but is non-trivial beyond one loop, since in $d\neq 4$ dimensions chiral symmetry is not conserved even for vanishing quark masses,
and might require an additional finite renormalization. The analogous cases of the axial and vector currents and scalar and pseudoscalar densities are discussed in Refs. \cite{Larin:1993tq,Collins}. 
In lattice schemes such as regularization-independent momentum subtraction schemes (RI-MOM), renormalization conditions  need to be imposed so that 
the Ward identities for the qCEDM and qCMDM are respected  \cite{Sturm:2009kb,Bhattacharya:2015rsa}.
Furthermore, the renormalization of these operators will involve some care as they involve an external gluon~\cite{Bhattacharya:2015rsa}.

In order to get a precise determination of the CPV couplings via our proposed relations Eq.~\eqref{eq:g0g1_mod} (with reasonable statistics), the lattice QCD calculations need be performed on the same ensembles.  
This will ensure the stochastic fluctuations are correlated allowing for a precise determination of the difference.
There are also dynamical sea-quark contributions to the matrix elements needed in these relations Eq.~\eqref{eq:fh_matrix_elements}, at least for the isoscalar qCMDM operators.
However, these contributions are suppressed in the large-$N_c$ expansion as at least two gluons are needed to connect the disconnected quark loop to the valence quarks.  There is an additional color trace for the closed quark loop, leading to a $1/N_c$ suppression compared to the connected contribution.
In practice, disconnected diagrams are found to be more suppressed than suggested from large-$N_c$ counting.
Therefore, for a determination of these couplings at the 20\% level, it is most likely safe to ignore the disconnected contributions. 
This expectation should be carefully checked in the numerical calculations.
A reliable estimate can be made with quarks that are heavier than the physical ones, in which the calculations will be numerically less expensive.

%
\section{Discussion}  
The main result of this work is the derivation of reliable relations that can be used to simplify lattice calculations of the CPV pion-nucleon vertices that originate in the up, down, and strange qCEDMs. 
By using chiral symmetry arguments the required three-point functions can be linked to spectroscopic two-point functions which are more suitable for lattice calculations. 
Similar relations had already been derived before, but these were found to be highly unstable under higher-order corrections \cite{Seng:2016pfd}. 
We have demonstrated that the modified relations in Eq.~\eqref{eq:g0g1_mod} linking the CPV couplings $\bar g_{0,1}$ to the up and down qCEDMs are preserved by all NLO and most of the N${}^2$LO corrections including effects from dynamical strange quarks. 
The remaining N${}^2$LO corrections have been conservatively estimated at $\mathcal O(20\%)$. 
We found that the strange qCEDM only induces leading contributions to $\bar g_1$, which are proportional to the numerically small $\pi^0$-$\eta$ mixing angle. 
As such, the relation involving the strange qCEDM identified in Eq.~\eqref{modStrange2} can obtain significant N${}^2$LO corrections. 

The lattice calculations we propose to extract accurate values for $\bar g_{0,1}$ are crucial to fully benefit from the impressive progress in the experimental EDM program. 
In heavy systems such as Hg, Xe, and Ra, the theoretical interpretation of the EDM limits (and hopefully future signals) suffers from large uncertainties that can be roughly divided into $50\%$ hadronic uncertainties (values of the LECs in terms of the quark-gluon CPV sources) and $50\%$ nuclear uncertainties (due to the complex nuclear many-body calculations). 
The strategy outlined here could go a long way in reducing the hadronic uncertainties. 
Furthermore, EDMs of light nuclei essentially only suffer from hadronic uncertainties \cite{Bsaisou:2014zwa,Yamanaka:2015qfa} such that this program would benefit even more from improved calculations of $\bar g_{0,1}$. 
The influence of improved matrix elements has been highlighted in several recent works. 
For instance, Ref.~\cite{Inoue:2014nva}  studied the impact of hadronic uncertainties on tests of electroweak baryogenesis scenarios finding that improved values of $\bar g_{0,1}$ would allow for more stringent tests. 
Refs.~\cite{Chien:2015xha, Cirigliano:2016njn} highlighted the role of hadronic uncertainties on constraining CP violation in anomalous interactions involving top quarks and Higgs bosons, finding that improved matrix elements could have as much impact in scrutinizing such couplings as new experiments. 
In summary, the recent impressive experimental progress must go hand-in-hand with theoretical developments. The strategy outlined in this work provides an important step towards such developments. 

\vspace{0.5cm}

\noindent {\bf Acknowledgements} -- 
We thank V. Cirigliano and T. Bhattacharya for several fruitful discussions.
EM  acknowledges support by the US DOE Office of Nuclear Physics and by the LDRD program at Los Alamos National Laboratory.
 JdV  acknowledges  support by the Dutch Organization for Scientific Research (NWO) 
through a VENI grant. The work of CYS was supported in part by National Natural Science Foundation of China under Grant  No.11575110 and No. 11175115, Natural Science Foundation of Shanghai under Grant  No. 15DZ2272100 and No. 15ZR1423100.
The work of AWL was supported in part by the U.S. Department of Energy, Office of Science: Office of Advanced Scientific Computing Research, Scientific Discovery through Advanced Computing (SciDAC) program under Award Number KB0301052; 
the Office of Nuclear Physics under Contract number DE-AC02-05CH11231;
the Office of Nuclear Physics Double-Beta Decay Topical Collaboration under Contract number DE-SC0015376
and the DOE Early Career Research Program under FWP Number NQCDAWL.

\bibliographystyle{elsarticle-num} 
\bibliography{su3}

\end{document}